\documentclass[manuscript,screen]{acmart}
\AtBeginDocument{%
  \providecommand\BibTeX{{%
    \normalfont B\kern-0.5em{\scshape i\kern-0.25em b}\kern-0.8em\TeX}}}

\setcopyright{rightsretained}
\acmConference[STT '23 at CHI '23]{April 23, 2023}{Hamburg, Germany}
\begin{document}

\title{A Multisensory Approach to Virtual Reality Stress Reduction}

\author{Rachel Masters}
\email{ramast1@colostate.edu}
\orcid{0000-0002-3857-6537}
\affiliation{%
  \institution{Colorado State University}
  \streetaddress{1100 Centre Ave}
  \city{Fort Collins}
  \state{Colorado}
  \country{USA}
  \postcode{80521}
}

\author{Francisco Ortega}
\email{f.ortega@colostate.edu}
\orcid{0000-0002-2449-3802}
\affiliation{%
  \institution{Colorado State University}
  \streetaddress{1100 Centre Ave}
  \city{Fort Collins}
  \state{Colorado}
  \country{USA}
  \postcode{80521}
}

\author{Victoria Interrante}
\email{interran@umn.edu}
\orcid{0000-0002-3313-6663}
\affiliation{%
  \institution{University of Minnesota}
  \streetaddress{200 Union St SE}
  \city{Minneapolis}
  \state{Minnesota}
  \country{USA}
  \postcode{55455}
}

\renewcommand{\shortauthors}{Masters, et al.}

\begin{abstract}
    Forest bathing is a nature immersion practice that reduces stress, restores mental resources, and has a wide variety of use cases in the treatment of mental illnesses. Since many people who need the benefits of forest bathing have little access to nature, virtual reality (VR) is being explored as a tool for delivering accessible immersive nature experiences via virtual nature environments (VNE's). Research on VNE's mainly utilizes the audiovisual capabilities of VR, but since forest bathing is a fully multisensory experience, further investigations into the integration of other sensory technologies, namely smell and temperature, are essential for the future of VNE research.
\end{abstract}




% A "teaser" image appears between the author and affiliation
% information and the body of the document, and typically spans the
% page.
\begin{teaserfigure}
  \includegraphics[width=\textwidth]{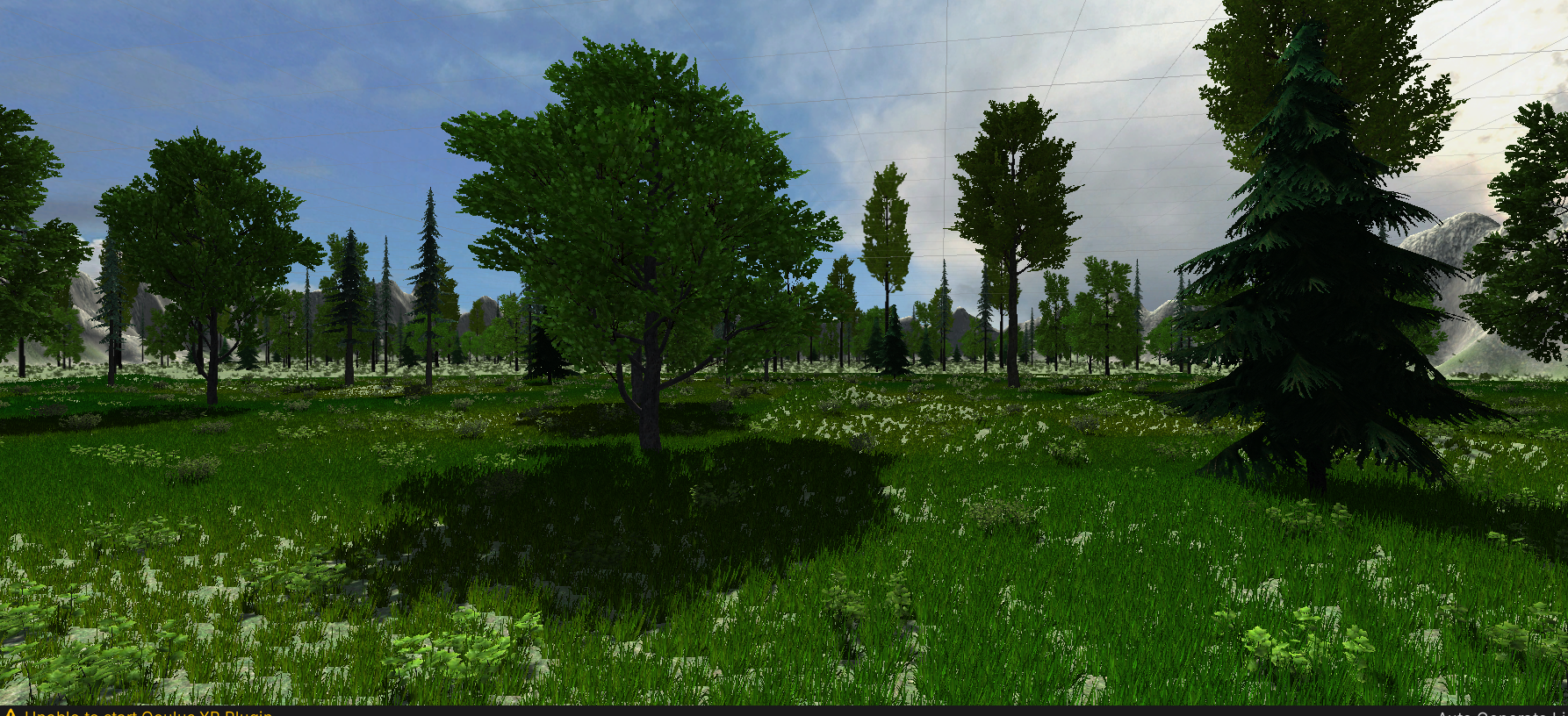}
  \caption{Multisensory VR Forest Bathing as Part of Addiction Recovery Application}
  \Description{Example of forest bathing experience for specialist guided addiction recovery}
  \label{fig:teaser}
\end{teaserfigure}

\maketitle

\section{Introduction}
Forest Bathing is a practice that reduces stress and restores mental resources via immersion in nature~\cite{park2010physiological}. While the practice is highly effective, many people lack nature access and therefore the ability to receive the restorative benefits of nature. Furthermore, the populations that lack access to nature are often the populations that need it most. For example, people in large cities often have less access to nature, yet the city lifestyle can be conducive to chronic stress~\cite{10665-345751}. Similarly, people in hospitals lack nature access yet could benefit from an immersive nature experience, as nature imagery in hospital rooms has already shown potential for improving patient pain~\cite{vincent2010effects}. While access to real nature is not always feasible, it is possible to provide similar, immersive nature experiences using carefully designed virtual reality (VR) simulations~\cite{9417782}. Since forest bathing is an immersive practice, VR is the perfect tool for replicating that nature immersion. 

Currently, new research in VR forest bathing focuses on the necessary qualities of a virtual nature environment (VNE) for optimal immersion and restorative impact. Modalities including screens of different sizes~\cite{DEKORT2006309}, 360 video~\cite{hedblom2019reduction}, and environments with 3D virtual assets~\cite{masters2022virtual} have been researched. Additionally, an active area of research involves the components of the VNE's themselves, including topics such as the necessity of biomass, or living green nature~\cite{masters2022virtual}, and of biodiversity~\cite{marselle2021pathways}, on restorative benefit. The most complicated and interesting research pathway, though, is on the subject of multisensory experiences. Most of this literature has been conducted on combining sound~\cite{conniff2016methodological,franco2017review}, and newer research has addressed the topic of smell~\cite{hedblom2019reduction}. 

Forest bathing is connected to the biophilia hypothesis~\cite{kellert1993biophilia}, which describes the biological need for nature and sensations experienced in nature. In MacKenzie's book on Human Computer Interaction, he mentions that beyond the five senses, humans have many other senses that shape the way that they perceive their surroundings~\cite{mackenzie2012human}. Forest bathing immerses people into nature via a multitude of senses including the five senses and others like pain and temperature. In urban forest parks, researchers are already investigating ways to fully include the five senses and beyond into urban park design for the purpose of creating a complex, cohesive sensory experience modelling true forest bathing~\cite{xu2022multi}. For urban landscapes, this is particularly challenging due to the noise and air pollution that exists in cities. For immersive, multisensory implementations using VR, the problem is even more complex and calls for further research leveraging new sensory technologies.

\section{Challenges}
A wide variety of challenges, considerations, and unknowns exist when designing multisensory VR nature experiences. For example, VNE's with sound have demonstrated restorative potential, but VNE's without sound have shown potential negative connotations involving sensation of predator presence~\cite{franco2017review}. Additionally, when adding sound to a virtual nature environment, many questions need to be asked in order to design sound that is plausible and fits with what people expect, or else immersion and presence will be damaged~\cite{slater1994depth}, and people may even experience cybersickness due to being in an environment that does not align with sensory expectations~\cite{jerald2015vr}.

Sound is easier to implement than smell because the VR headsets already have integrated sound capabilities. Smell presents more challenges because in addition to asking the same questions asked of sound design, smell design involves the integration of external odor interfaces into the existing VR interface. Questions about how to create odor that fits a given type of nature environment, how to alter odors like soundscapes for different environments, how to build complex odors that reflect the different elements in a nature environment, and how to understand the way people interact with odor as a component of nature all remain unanswered questions that call for more extensive future research. Also, ways to introduce the smell interfaces themselves while maintaining immersion in virtual environments is another challenge faced, considering factors like weight, position, and contact with the smell technology which have all been little researched in the context of VNE's. Other senses beyond the five senses, like temperature, wind, pain, and sensed air pollution are challenging to observe and complex in similar ways as smell, yet they are equally as interesting and little researched for an immersive virtual nature experience.

\section{Position: Call for Multisensory VNE's Using Smell and Temperature}
Existing research suggests that both smell and temperature can enhance the sense of presence in a VNE. Smell engagement has been shown to have significant effects on restorativeness in VNE's compared to VNE's with no odor ~\cite{hedblom2019reduction}. Smell has also been shown to have significant effects on the sense of presence in a virtual environment, and can increase attention to the environment~\cite{jiang2016effect, munyan2016olfactory, munyan2016vr}. Temperature has been shown to improve perception of an environment~\cite{chen2017thermally}, and tactile cues including temperature and wind have had a positive impact on experience and presence~\cite{hulsmann2014wind}. The existing body of literature shows promise for future work related to the full integration of smell and temperature cues into immersive VNE's. Our position is that it is important to consider how the challenges of integrating smell and temperature can be overcome to enhance the effectiveness of VR forest bathing experiences. This will involve a combination of research on fitting smell and temperature technologies to audiovisual VNE's, research on integrating the technologies into our systems in a seamless way, and research testing if the technology is working as intended.

Research projects on scent will relate to pinpointing and understanding complex scents present in forest environments, simulating those scents using olfactory technology, integrating that technology with fitting VNE's, and observing the additional benefits of using tailored scent technology across diverse populations. Research on temperature technologies will involve understanding what expected temperatures are for different areas of a natural environment, simulating temperature as expected in different areas of a VNE, and measuring additional restorative benefit provided by improving perception of environments using temperature cues. Additionally, research will explore the balance needed between audiovisual, olfactory, and temperature stimuli such that people have an improved experience without sensory or cognitive overloading. Many more research projects can and should be conducted with respect to smell, temperature, and other senses. The completion of the initial projects proposed in this section will ultimately result in developing a open-source framework for creating optimal and replicable, multisensory VNE's that can be applied to many different use cases in the future, a few of which are listed below in Section~\ref{usecase}.

\begin{figure}[ht]
\centering
\begin{minipage}{.5\textwidth}
  \centering
  \includegraphics[width=1\textwidth]{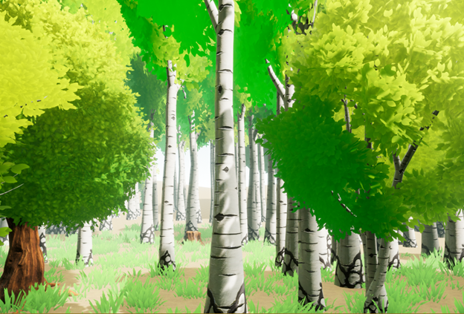}
\end{minipage}%
\begin{minipage}{.5\textwidth}
  \centering
  \includegraphics[width=1\textwidth]{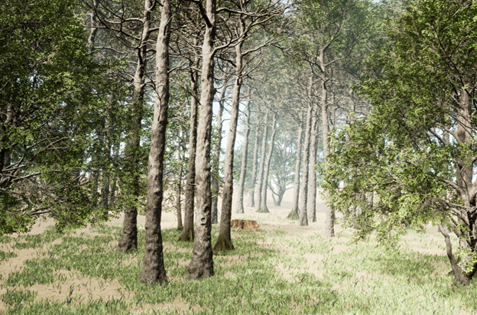}
\end{minipage}
\captionof{figure}{Use Case for Smell Enhancing Lower and Higher Realism Environments}
\label{fig:fig2}
\end{figure}

\section{Use Cases} \label{usecase}
The use cases of smell and temperature technology for increasing immersivity in VNE's are many and varied. A prime use case is improving easily deployable VNE's. One large challenge in the design of VNE's is creating an effective, everyday product that is deployable on portable platforms like the Oculus Quest 2. In order to create an immersive VNE, the VNE has to have enough interesting qualities and interaction to keep the user engaged~\cite{jerald2015vr}. Often times, this includes the placement of many diverse, highly realistic assets that currently require state of the art machines to render. Deploying one of these environments on the Oculus Quest 2 would result in choppy movement and acceleration that is cybersickness inducing~\cite{jerald2015vr}. In our current research, we have been investigating if a high level of detail of the virtual plants is critical for the immersive quality, or if a lower realism, more artistic yet also aesthetic rendering of a nature environment can be restorative. Figure ~\ref{fig:fig2} illustrates example environments, one of high realism and one of lower realism. The lower realism would be easier to render with movement and exploring, which would increase interest and presence in the environment. Smell interfaces would introduce a whole new layer of possibilities onto this research with respect to overcoming the audiovisual limitations of existing VR technology. Research could be conducted on whether smell design can make lower realism environments more compelling. If the implementation of smell is successful, a highly immersive, everyday stress relief experience could be created and deployed on portable hardware like the Oculus Quest 2, and the accessibility of VR stress reduction benefits would increase despite the optical limitations of existing hardware.

Another increasingly more pertinent application of highly immersive VR forest bathing experiences is related to mental illness and health. In a longitudinal study on forest therapy, smells were found to help people with stress induced mental illness recover via triggering positive associations and increasing awareness~\cite{palsdottir2021garden}. Understanding how to effectively integrate olfactory stimuli in VNE's can improve the technology to help address mental illness and bring solutions to people who lack regular or easy nature access. Lately, forest bathing has also been suggested as an aid for addiction recovery~\cite{kotera2020commentary}. Since forest bathing involves heightened awareness of one's senses, it has the potential to be coupled with mindfulness based (MB) therapy techniques to create more powerful recovery solutions~\cite{garland2014mindfulness}. In another research project, we have been investigating ways to couple professionally guided, MB addiction therapy and forest bathing in VR to increase accessibility to therapy for rural youths struggling with addiction. Figure ~\ref{fig:teaser} shows an example of what a forest bathing experience for this context looks like. Incorporating carefully crafted smell and temperatures into this environment would enhance the sensory engagement, user awareness and presence, and therefore the efficacy of the treatment in this environment, all while maintaining deployability on an accessible device like the Oculus Quest.

\section{Conclusion}
VR forest bathing is becoming an increasingly more researched, interdisciplinary tool with many use cases. However, in order to mimic real forest bathing, VR must offer a truly immersive yet accessible experience. As we fight to overcome audiovisual limitations and enhance experiences beyond the existing audiovisual landscape, incorporating the other senses becomes a pertinent avenue of research. The future of the field is moving towards leveraging new technologies to create greater senses of presence, and smell and temperature technologies are critical for this cause.

\section{Acknowledgements}
We would like to acknowledge the National Science Foundation for
grant numbers: 2238313, 2223432, 2223459, 2106590, 2016714, 2037417, 1948254.

\bibliographystyle{ACM-Reference-Format}
\bibliography{sample-base}

\appendix

\end{document}